\documentclass[twocolumn,showpacs,prl]{revtex4}
\usepackage{graphicx}
\usepackage{dcolumn}
\usepackage{bm}
\begin{document}

\title{\bf  An Argument for Nonminimal Higgs Coupling to Gravity }
\author{Y. N. Srivastava}
\affiliation{INFN \& Dipartimento di Fisica, Universit\'{a} di Perugia, Perugia, Italy}
\author{J. Swain and A. Widom}
\affiliation{Physics Department, Northeastern University, Boston, MA 02115 USA }

\begin{abstract}
The coupling of gravity to a scalar field raises a number of interesting questions
of principle since the usual minimal coupling obtained by replacing ordinary
derivatives with covariant derivatives is not available -- they are the same
operation on scalar fields. Conformal couplings in the Lagrangian proportional 
to $\phi^2 R$ have been suggested before, usually to maintain conformal invariance for 
massless scalar fields, but at the cost of breaking the equivalence principle.
Here we give intuitive arguments for the appearance of such a term due to
fluctuations of scalar particles about their classical world lines. Remarkably,
these arguments give precisely a correction of the form required to maintain
conformal invariance. We also show that such a term would naturally be
expected for the Higgs field in the Standard Model, making a perhaps surprising
connection between weak-scale physics and gravity. The nonminimal coupling,
whether induced by quantum corrections or already present as a bare term, 
can be constrained from measurements of the Higgs width assuming the Higgs 
particle is to be detected.
\end{abstract}

\maketitle

\section{Introduction}

The coupling of scalar fields to gravity is a matter of some subtlety. The usual
minimal coupling prescription obtained from a flat spacetime Lagrangian by
replacing ordinary derivatives with covariant ones is not available since
both derivatives are the same when acting once on scalar fields. The associated geometrical
picture is quite clear. Curvature at a point $x$ is measured classically by parallel-transporting
an object with a Lorentz (e.g. a vector) index around an infinitesimal closed path, say
the parallelogram spanned by $dx^{\mu}$ and $dx^{\nu}$ in a tangent plane at $x$. 
The curvature tensor $R^\alpha_{\beta\mu\nu}$
associates an infinitesimal rotation matrix $R^\alpha_{\beta\mu\nu}dx^\mu \wedge dx^\nu$ to that plane which
represents the fact that the object comes back to $x$ rotated. A scalar test particle,
classically at least, will not have any way of indicated how much it has been rotated,
so it cannot detect (or couple to) curvature via a covariant derivative.

For a scalar field $\phi$ a nonminimal coupling 
$\xi\phi^2 R$ is often introduced via discussions of the conformal anomaly \cite{QFT_curved_spacetime}.
Within this picture one starts with a massless scalar field, which is conformally invariant,
and finds that a coupling of the form $\xi\phi^2R$ is required so that the trace of the stress-energy
tensor $T^\mu_\mu$ is zero. For a general scalar field one expects $T^\mu_\mu$ should
be zero when the mass of the corresponding particle is zero. 

For a $(+++-)$ signature metric, a massive scalar field then obeys the equation:
\begin{equation}
-\hbar^2 D^\mu D_\mu \phi +m^2c^2\phi +  \xi \hbar^2 R\phi=0, 
\label{scalar1}
\end{equation}
which can be derived from a Lagrangian density
\begin{equation}
{\cal L}= \frac{1}{2} 
\left( -\hbar^2 g^{\mu \nu} \partial_\mu \phi \partial_\nu \phi 
-m^2c^2 \phi^2 -\xi \hbar^2 R\phi^2 \right).
\label{scalar2}
\end{equation}
which has $T^\mu_\mu \to 0$ as $m\to 0$ if $\xi=1/6$.
While there has been some debate in the
literature (see especially the last reference cited in \cite{QFT_curved_spacetime}) over the years, it
seems generally accepted that for $T^\mu_\mu=0$ and  conformal invariance to be maintained,
one must choose $\xi=1/6$.

An alternative point of view, which we take here, is to calculate what quantum corrections
one would expect to the Klein-Gordon equation for a massless scalar particle. Here we
present a heuristic derivation which we feel has the virtue that it makes the nature of
the corrections and the naturalness of a non-mininimal coupling quite intuitive.

Start with the massless Klein-Gordon equation in momentum space in obvious notation:
\begin{equation}
[E^2-c^2(p_x^2+p_y^2+p_z^2)]\phi=0. 
\end{equation}
On shell we can (even if only locally, but this is a local equation) choose the
particle to go along the $z$ axis so that $E^2-c^2p_z^2=0$ and $p_x=p_y=0$.
This is the classical path about which we want to find fluctuations in a space
with some curvature.

To leading order in curvature (and without derivative couplings to 
curvature) we can take the curvature to be represented by a 
scalar 4-curvature $R$ which is not zero. Any other terms proportional
to the curvature tensor with uncontracted indices (like the Ricci tensor)
are ruled out by general covariance since $\phi$ is a scalar.

Now consider the particle path fluctuations in the $x$ and $y$ directions
(the directions which are off the classical path and can be modelled as small
perturbations -- we discard the rather drastic large fluctuation from the 
$x=ct$ path to the $x=-ct$ path which we would effectively have as antiparticles
anyway). Note that fluctuations off the classical path are fluctuations off
the light cone, so one might well expect a breaking of naive conformal
invariance associated with the particle classically following light-like null
paths only.

A slice in the x-y plane through the 4-sphere of radius $r$ has curvature $1/r$
as does one through the x-y plane. Associate a momentum due to being
localized by that curvature of $p=\hbar/r$. A more rigorous derivation will
be presented in a later paper\cite{later_nonmin_Higgs}.
Going back to the Klein-Gordon equation we now have:
\begin{equation}
[(E/c)^2-p_z^2-(\hbar/r)^2-(\hbar/r)^2)\phi=0
\end{equation}
or, restoring the fact that the $z-axis$ was arbitrary, 
\begin{equation}
[(E/c)^2-p_z^2-2 (\hbar/r)^2 ] \phi=0.
\label{energy1}
\end{equation}
Now we want to connect $r$ to the scalar curvature $R$.

$R$ is the sum of sectional curvatures in each plane with each orientation counted
separately\cite{Frankel}. For example, in 2 dimensions, $R=K(x\wedge y)+K(y\wedge x)=2/r^2$ which gives
the usual factor of 2 from elementary differential geometry. 
Here we have 12 planes
(6 for the usual boost and rotation planes times 2 for 2 orientations) so that
$12/r^2=R$ or $1/r^2=R/12$. Using Eq.(\ref{energy1}) 
\begin{equation}
\left[(E/c)^2-p_z^2-\frac{R}{6}\hbar^2 \right] \phi=0
\end{equation}
or, in general
\begin{equation}
\left(-\hbar^2 D^\mu D_\mu + \frac{1}{6}\hbar^2R \right) \phi=0
\end{equation}
Remarkably with this heuristic argument, the $1/6$ factor is natural and the physical interpretation
of the result is clear -- the extra ``mass'' terms are due to an effective potential due to curvature
that acts as a mass. It is naturally of order $\hbar^2$ and proportional to $R$.

This then describes quantum corrections to the wave equation of a particle
that's massless in flat spacetime (which is the only place that statement really
makes sense since mass is the value of a Poincare group Casimir so it really
{\em assumes} flat spacetime). Intuitively,  a quantum scalar particle is able to
measure curvature without having a Lorentz index since it feels out the curvature
by its fluctuations -- this has no classical analog. The fact that the particle is
not localized to a single path also explains the apparent violation of the classical
equivalence principle -- there is no sensible ``Einstein elevator'' of arbitrarily small
size.

\section{A nonzero bare $\xi$?}

Once one has acknowledged the fact that a term of the form $\xi R\phi^2$ in the Lagrangian
will generally be induced by quantum effects, it is by no means obvious that one can
rule out a bare term of almost any value. The term is in many ways analogous to a
magnetic moment term, involving as it does a direct coupling to a curvature -- here
to gravitational curvature, while in the electromagnetic case, to the curvature $F_{\mu\nu}$
of the $U(1)$ connection that describes electromagnetism. Nonminimal couplings for
electromagnetism are ruled out on the ground of renormalizability, but such arguments
are not available in theories with Einstein gravity. Feynman, in his lectures
on gravitation\cite{Feynman} discusses such a coupling and points out that the classical
value is essentially arbitrary. Since we have not yet had any experimental
access to any fundamental scalar fields, it is interesting to ask what such a term
would apply for the Higgs field of the Standard Model.

\section{The Higgs boson}

In the Standard Model, the scalar field used to give masses to particles is
actually, before the field takes its vacuum expectation value, a massless
scalar field with a quartic self-coupling. Neglecting coupling to other
fields, but including the nonminimal scalar coupling to gravity, we have

\begin{equation}
{\cal L}_{Higgs}=-g^{\mu \nu}
\partial_\mu\phi^\dag\partial_\nu\phi - \lambda \left(\phi^\dag\phi-\frac{v^2}{2}\right)^2 - \xi\phi^\dag\phi R.
\end{equation}
where $v$ is its vacuum expectation value. Making the usual expansion about $\phi=\frac{v+H}{\sqrt{2}}$ where 
$H$ is the experimentally detectable Higgs field,
and expanding the term $\xi R\phi^\dag\phi$ then gives the terms:
\begin{equation}
 L_{gravity-Higgs}= -\frac{1}{2}\xi v^2R-\frac{1}{2}\xi H^2R-\xi v HR
 \end{equation}
The first term can be interpreted as a (finite) renormalization of Newton's constant,
the second as a correction to the mass of the physical Higgs particle, and
the last term as a coupling of the Higgs boson to gravitons which contributes
to the Higgs decay width.

We leave a discussion of the effects of fluctuations due to other fields
on the parameter $\xi$ to
another paper\cite{later_nonmin_Higgs}. Note that if for any reason one believes that $R$ is in fact large
(for example due to rolled up extra dimensions which would contribute large
sectional curvatures to $R$) then $\xi$ could become very large as well. There are
also clearly interesting contributions of nonminimal couplings of the kind described
here to cosmological models involving scalars, a topic to which we plan to return
in a later publication.


\thebibliography{apssamp}
\bibitem{QFT_curved_spacetime} See, for example, N. D. Birrell and P.C. W. Davies,
``Quantum Fields in Curved Space'' (Cambridge University Press, 1982); S. A. Fulling, ``Aspects of Quantum Field Theory in
Curved Space-Time'', London Mathematical Society Student Texts {\bf 17} (Cambridge University Press, 1989);
R. M. Wald, ``Quantum Field Theory in Curved Spacetime and Black Hole Thermodynamics'' (The University of Chicago Press, 1994);
L. Parker and D. Toms, ``Quantum Field Theory in Curved Spacetime'' (Cambridge University Press, 2009);
 F. Bastianelli and P. van Nieuwenhuizen, ``Path Integrals and Anomalies in Curved Space'' (Cambridge University Press, 2009).
\bibitem{Frankel} See, for example, T. Frankel, ``The Geometry of Physics'' (Cambridge University Press, 1997).
\bibitem{Feynman} R. P. Feynman {\em et al.}, ``Feynman Lectures On Gravitation''
(Frontiers in Physics, 2002).
\bibitem{later_nonmin_Higgs} in preparation
\end{document}